# Integrating Travel Demand and Network Modelling: a Myth or Future of Transport Modelling

Ali Najmi[a], David Rey[a], Taha H. Rashidi[a1], S. Travis Waller[a]

[a] *Research Center for Integrated Transport Innovation, School of Civil and Environmental Engineering, The University of New South Wales, Sydney, Australia, 2032*
(a.najmi@unsw.edu.au, d.rey@unsw.edu.au, rashidi@unsw.edu.au, s.waller@unsw.edu.au)

In this paper, a novel transport planning model system (TPMS) is formulated which is built on the concepts of supernetworks, multi-modality, integrity and calibration. In the proposed formulation, activity travel pattern (ATP) choice facets including the choices of activity, activity sequence, mode, departure time, and parking location, are all unified into a time-dependent supernetwork. The proposed model accounts for the dynamicity of the network, including time-of-day and congestion effects. These help capturing the interdependencies among all different attributes of a full transport planning system. Moreover, the proposed TPMS explicitly formulates an operating capacitated public transport system. To allow visiting locations multiple times and to alleviate the complexity of the proposed supernetwork, a novel multi-visit vehicle routing problem is proposed which does not enumerate the node and link visits. In order to calibrate the model based on the major travel attributes of the travel survey data, a set of splitting ratios are introduced to distribute trips on the supernetwork. The model uses the splitting ratios to integrate the supernetwork and the traffic assignment model in a unified TPMS structure. At last, numerical examples are provided to demonstrate the advantages of the proposed approach.

## 1. Introduction

Research on Transport Planning Model Systems (TPMSs) such as activity-based models (ABMs) has recently attracted a lot of attention. TPMSs include a number of sub-models such as trip generation, destination choice, mode choice, activity choice, activity chains, route choice, and traffic assignment models which are linked together. In the standard practice, the sub-models are partly developed individually, usually based on maximising an objective function which can be total travel time, or log-likelihood function. They are then connected in a sequential and ad-hoc manner so that the outputs of one or some models are fed into the other models. The fundamental limitations of such sequential approaches are well-discussed in Boyce (2002) and Najmi et al. (2018a,b).
There are two main limitations that make the sequential structure incapable of properly capturing the synchronisations among the sub-model outputs. First, due to the absence of spatiotemporal constraints in a physical network in the unconstrained econometric (dis-)utility minimisation/maximisation modelling approaches, the interaction between the models is lost (Jara-Díaz, 2003; Recker, 2001). Second, the optimization formulation resulting from the combination of traffic assignment models and demand-side models is non-convex in general, which makes the convergence of the integrated TPMSs slow and often impractical.

Asynchronisation among model components in conventional TPMSs such as trip-based models and ABMs unrealistically affects the activity travel patterns (ATPs) generation process of travellers. This is a limitation of conventional models which cannot fully capture the temporal and spatial dimensions of the entire problem. Specifically, the appropriate treatment of the temporal and spatial dimensions is perhaps the most important prerequisite to generate precise ATPs (Pinjari and Bhat, 2011). Therefore, the asynchronisation in addition to the needs of having detailed ATPs of travellers has triggered researchers to develop different unified demand-side models (or so-called ATPs generators) (e.g. Liao et al., 2010, 2013; Liu et al., 2015; Chow and Djavadian, 2015; Ouyang et al., 2011; Fu and Lam, 2014). Thus, *ATPs generator*s are unified formulations that can generate all the demand-side choice facets of travellers simultaneously and usually for a whole day. As in conventional models, spatiotemporal constraints have not sufficiently received attention in the ATPs generators; possibly due to their complexity. Nonetheless, there are some studies in which spatiotemporal constraints play a key role in model development (e.g. Chow, 2014; Chow and Djavadian, 2015). Owing to the importance of synchronisation in the precision of the TPMSs, we develop an integrated model which not only incorporates a unified ATPs generator at the presence of spatiotemporal constraints, but also attempts to synchronise the ATPs generator and traffic assignment model.

Furthermore, to obtain optimal ATPs, proper modelling of the interactions among different travel modes in the transport systems is necessary. The existing analytical equilibrium models of multimodal systems are solely based on trip-based demand (Chow and Djavadian, 2015). The trip-based structure disjoints the inter-modal connections among the models. This is also in contrast with the reality that demand for multimodal transport systems has a high correlation with activity schedules of travellers so that the availability or accessibility of a mode can remarkably change the activity agenda. This necessitates modelling the multimodal transport systems and activity schedules of travellers in a unified structure. Modelling the inter-modal interactions allows representing the multimodal dynamicity of the system and compromising between cost and time in the multimodal structure (Resat and Turkay, 2015). Specifically, the public transport timetable is another parameter that may affect ATPs for travellers. Many studies address the capacitated time-dependent public transport problem; however, these models are usually implemented on discrete space-time networks (Lu et al., 2016; Liu and Zhou, 2016; Liu et al., 2018). The models are usually network design problems which seek the answer of strategic decisions such as constructing new transit lines and adding or optimising train or bus schedules through a static origin-destination (OD) demand input (Liu and Zhou, 2016; Martínez et al., 2014). Nonetheless, at the microscopic level, studies on the interactions between the public transport, private transport and demand are limited. Thus, we attempt to address the multi-modality of the transport systems as well as the interaction of public transport and private vehicle modes.

The output of activity scheduling, whether having temporal and spatial dimensions or not, usually includes Origin-Destination (OD) matrices that should be loaded to a traffic assignment model. Nonetheless, the aggregate trips are not stable because if the resulted ODs are imported to the traffic assignment model, the resulted travel times and dis-utilities may be significantly different from the ones that originally were used for activity scheduling (Cools et al., 2010). The mismatch in interaction between demand-side and traffic assignment models is very common (Najmi et al., 2019) which usually leads to the most inordinate asynchronisation in TPMSs. This is mainly due to the highly non-convex solution space of the interacting models. The problem is more common for multimodal and dynamic TPMSs. To alleviate the discrepancy in the literature, demand-side and traffic assignment models are iteratively solved (at the presence of feedback loops) until convergence which is minimising the discrepancy between travel time estimations of consecutive iterations (e.g. MORPC: Parsons Brinckerhoff, 2005; ALBATROSS: Arentze and Timmermans, 2004a; NYMTC: Parsons Brinckerhoff, 2014). However, the feedback loops are present only at the simulation step and their impacts on the calibration parameters (in the calibration step) are not properly addressed in the literature (Najmi et al., 2019). In the proposed approach, we attempt to address the interaction not only within the simulation step but also within the calibration step of TPMSs.

Regardless of whether the feedback loops are considered in the calibration step or not, the calibration of the unified ATPs generators itself is challenging. Since the expanded network concept is usually used in unified ATPs generators, and the choices of nodes in the networks are not exclusive, the calibration of the unified ATPs generator to household travel survey data is very challenging and the conventional calibration approaches are not applicable in the models (see Chow and Recker, 2012). However, there are some efforts to calibrate the parameters of the unified models. In Recker et al. (2008) and Chow and Recker (2012), a genetic algorithm and an inverse optimisation approach have, respectively, been proposed to calibrate some parameters limited to the ATPs of households (household level properties). However, the main difference between their work and the current paper is that the collective behaviour of travellers (system-level properties) is not investigated in their work while it is a main contribution of our approach. System-level properties are of utmost importance in the structure of TPMS because it is the system level travelling behaviour that forms the traffic assignment model (Najmi et al., 2018a). Ignoring the interaction between the ATPs generators and traffic assignment models could result in significant stability issues, error propagation and transferability restrictions (Najmi et al., 2018b).

Furthermore, using the ATPs generators in the body of TPMSs, in the presence of multiple traffic assignment models, is an interesting topic that has not received enough attention (Najmi et al., 2019). The inconsistency in the scheduling period of ATPs generators, which is usually a whole day, and in the time period of traffic assignment models, each of which is usually for few hours, makes the formation of a TPMS complicated. One of the main reasons is that congestions in different time periods concurrently affect the daily ATPs of all travellers and vice versa. We attempt to address this complexity in this research.

- **Our contributions**

In this paper, we formulate a novel integrated TPMS structure in which a unified ATPs generator and multiple traffic assignment models are integrated. We make multiple theoretical contributions, each of which having specific practical implications. First, we provide a comprehensive formulation for a unified ATPs generator in which we attempt to incorporate the above-mentioned aspects of an ideal demand-side model such as spatiotemporal constraints, multi-modality, as well as private and public transport all together. The ATPs generator is integrated with a number of traffic assignment models corresponding to specific time periods of the entire planning period (typically a day) to capture transport demand-supply dynamics. Second, to simplify using expanded, discrete space-time networks, we embed a new multi-visit vehicle routing formulation in the proposed integrated TPMS which allows visiting nodes and edges of the network multiple-times throughout the planning period. Third, using a number of feedback loops, the proposed model iteratively and dynamically updates travellers' daily itinerary while accounting for dynamic travel times (i.e. congestion effects) at different time periods of the planning period until convergence. Fourth, a new calibration solution, using splitting ratios, is proposed to effectively calibrate the proposed TPMS as a whole. Not only do the splitting ratios control system-level properties of the transport system, they also take into account the interaction among travellers. Lastly, we provide ample numerical examples to illustrate the insights of the proposed approach, as well as numerical results. The analysis conducted reveals the critical role of splitting ratios for reproducing the observed travel patterns as well as in speeding up the convergence of the TPMS. The outcomes also highlight the critical role of feedback loops in the proposed model and in integrated ATPs generators in general.

## 2. Literature review

In this section, parallel research on the ATP generation is presented. Section 2.1 presents the standard practice of ATPs generation approaches that are widely used in commercial transport packages. Sections 2.2 and 2.3 illustrate the novel approaches in ATPs generation which are developed to circumvent problems in the conventional ATPs generators by unifying different model components. Lastly, Section 2.4 explores two approaches for linking demand and traffic assignment models.

### 2.1. Econometric, simulation, rule and heuristic, -based ATPs generators

TPMSs, in practice, usually cannot guarantee seeking the optimum ATP for each traveller as it is a computationally challenging task. Therefore, modellers are more inclined to estimate the feasible space–time region for travellers. Accordingly, ABMs usually deploy a series of models including 1) utility maximisation-based models (i.e., multinomial logit and nested logit models)(Sacramento: Bowman et al., 2006; CEMDAP: Bhat et al., 2004; FAMOS: Pendyala et al., 2005) and 2) rule-based models (such as TASHA: Roorda et al., 2008; ALBATROSS: Arentze and Timmermans, 2004a) to circumvent the feasible space–time complexity. Nonetheless, these models have been extensively used in practice and in developing large-scale models with their focus being on simulation of activity patterns which opposes the spatiotemporal constrained scheduling behaviour (Chow and Nurumbetova, 2015). Using these models, some travelling decisions (such as trip purpose and activity sequence) are initially made for each traveller and then, using the space-time constraints for fixed activities (such as work and school), the full ATPs of travellers is heuristically scheduled. There is a rich body of research under this category; however, we do not further illustrate the technical issues of these efforts because our focus is not on this category. Interested reader is referred to Pinjari and Bhat (2011).

### 2.2. Shortest path-based supernetworks

In line with the efforts to develop unified ATPs generators, a stream of expanded network-based (also known as supernetwork) models has been introduced in which the ATP for a traveller can be obtained by running classic shortest path algorithms onto an expanded network. The concept of supernetwork was first introduced by Daganzo and Sheffi (1977) to represent a multi-modal transport network. Their proposed representation has been the main building block of the supernetwork research. In the representation, to interconnect different single-modal networks, transfer links which connect the modal networks at the same physical locations were added. Following this idea, a supernetwork can be constructed with connecting many independent networks each of which for an individual mode-time period. Later, Nagurney and Dong (2002) introduced transaction links to model activity implementation. In their representation, location choice and route choice can be modelled simultaneously. Although any path through their proposed expanded network represents route choice, multi-modal networks were not taken into account.

In the last 15 years, the so-called supernetwork models have been developed to address the integrated structure of transport systems. Arentze and Timmermans (2004b) suggested multi-state supernetworks, which unified activity programs of travellers, multi-modal transport networks and locations of activities. Their supernetworks were constructed for each traveller separately and are composed of as many copies of physical networks as different modes and activity states in an activity program execution. In their approach, while the travel mode state determines which particular mode is used, if any, the activity states determine which activities have been conducted. Later, this formulation is extended by a stream of research at Eindhoven University. As the developed supernetworks in Arentze and Timmermans (2004b) became very large and possibly intractable even for a small activity program, Liao et al. (2010) used separate sets of private vehicle networks, public transport network and transition and transaction links to scale down the size of the supernetwork. To further reduce the size of the network, Liao et al. (2011) proposed a heuristic approach to select small set of locations for each traveller. Later, they added a temporal dimension to their suggested supernetwork by incorporating time-space constraint in their formulation (Liao et al., 2013). Ramadurai and Ukkusuri (2010) proposed another unified framework, referred to as activity-travel networks, to model activity location, route choices, and activity duration, simultaneously using a supernetwork representation of the problem subject to a dynamic user equilibrium condition. The authors assumed an aggregate traffic assignment to measure congestion; however, they omitted the details of the scheduling constraints that are specific to each decision maker (Chow, 2014).

While the literature on supernetwork-based representation of a physical network and activities is rich, the time-dependent activity-travel assignment models of supernetworks are limited and not well developed (Liu et al., 2015), because time dimension significantly increases the size of supernetwork. Ouyang et al. (2011) proposed a model for solving the daily ATP scheduling problem by constructing an expanded time–space network which is extended by Fu and Lam (2014) to include uncertainty in the network. Liu et al. (2015) proposed a formulation, so-called dynamic activity-travel assignment (DATA), that is a discrete-time dynamic user equilibrium (DUE) model, in which any path through a personalised supernetwork represents an activity-travel pattern at a high level of spatial and temporal detail. In Liao et al. (2013), space-time considerations are incorporated in the supernetwork formulation, however, , it is limited to space-time considerations in selection (filtration) of location sets to be included in the expanded network.

There are some critical problems with the reviewed models. In contrast to econometric-based models, the temporal dimension has been used in supernetworks; however, it is limited to time-discretised supernetworks which remarkably increase the network size. Furthermore, the structure of the developed models in the literature is a concatenation of selected locations and connections distributed at different activity-vehicle states. This structure comes with an explosion of the network scale and as a result, the optimisation on the network is intractable even for a small number of activity nodes (Liao et al., 2013). Despite the fact that some heuristics have been used to scale down the size of the personalised network by choosing a small set of locations, it is not an easy task as the (dis-) utility of choosing a location not only does depend on the (dis-) utility of that location, but also depends on the sequence of activities, the activity duration, departure time, and most importantly the network condition. Thus, scaling down the expanded network is a risky activity as it may affect the validity of the model.

## 2.3. ARP-based models

To optimally schedule the activity-travel pattern of travellers, activity routing problem (ARP) formulations have been widely used in the literature. It should be mentioned that ARP and vehicle routing problem (VRP) are identical; therefore, for the sake of simplicity, we use the ARP term to explain the concepts throughout the paper. Recker (1995) for the first time mathematically formulated an activity routing problem, named Household Activity Pattern Problem (HAPP) by proposing a pickup and delivery optimisation models for the scheduling the household activity patterns in which the space-time constraint is incorporated; nonetheless, it suffers from restrictions on the choice dimensions covered such as route, mode and parking. After that, to embed a destination choice model into the routing and scheduling considerations of daily activities, the generalised ARP was introduced which allowed selection of a single location for conducting an activity among a number of candidate locations (Kang and Recker, 2013; Chow and Liu, 2012). Later, in Chow and Djavadian (2015), the multimodality was added to the original version of HAPP to improve its functionality.

Distinct from the majority of activity-based travel demand modelling techniques pointed out in the previous section, ARP-based models can offer spatiotemporal constraints as the space–time prism is associated with each activity and each traveller. Continuity of time in these models results in a relatively smaller expanded network time; nonetheless, the solution algorithms for the models are much more computationally demanding than those for the time-discretised supernetworks-based models (VRP versus shortest path algorithms).

## 2.4. Linkage of demand (ATPs) and traffic assignment model

The interaction between travel demand and traffic assignment models has been the backbone of the contention among researchers (Najmi et al., 2018a; Lin et al., 2008) which is mainly due to the incompatibility between the outputs of the models. Demand outputs that are consequences of loading travel times into demand models, if loaded into the traffic assignment model, may produce updated travel times that are different from the initially loaded travel times to produce the demand outputs. This incompatibility has triggered three main research approaches for dealing with the interaction between the demand-side and traffic assignment models.

In the first category, the main purpose is running microscopic and mesoscopic analysis on the transport network so, the OD matrices, as the representatives of demand models, are linked to traffic assignment models. As loading the OD matrices usually does not result in observed traffic counts statistics, it is very common to update the OD matrices using OD calibration methods (e.g. Spiess (1987), Cascetta and Nguyen (1988), and Lundgren and Peterson (2008)). Although the updated OD matrices are usually fully compatible with the traffic assignment model, there is actually no linkage between the original demand-side models and traffic assignment model. Neither the demand-side models are available to update the OD matrices, nor is feedback loop to transfer the network congestion to the demand-side models. This category is not under the scope of this study so the interested reader is referred to Najmi et al. (2018a) for further information.

In the second category, some researchers have tried to develop combined models to fully remove the incompatibly in the models. In this regard, behavioural assumptions are translated into mathematical conditions, and then seek approximate solutions that satisfy these conditions (Bar-Gera and Boyce, 2003). In these models, travellers are often divided into some classes, either by socio-economic attributes or the purpose of their travel, assuming that travel-decision characteristics are the same within each class (e.g. same value of time and similar sensitivity to travel times in choosing their origin, destination and mode of travel) but differ among classes (Boyce and Bar-gera, 2004). Thus, the combined models are aggregates in all the perspectives which are applicable in macro and meso-level forecasting of travel demand. For a comprehensive review of the combined models, the interested reader is referred to Boyce (2002). The major problem with the combined models is that they suffer from lack of detailed ATP for each traveller.

In the third category, the focus is on simulating the travel demand and detailed ATPs. Also, there are some efforts to consider the impact of the network congestion in the scheduling process. Accordingly, after scheduling the travel demands over time, the scheduled trips are usually imported into the traffic assignment models (in form of ODs) for network-side analysis. Using feedback loop, the updated travel times (congestion) are fed back to the demand-side models to update travel demands and ATPs based on the measured congestion of the network. Despite the fact that these efforts connect demand-side models to traffic assignment models (Konduri et al., 2011; Hao et al., 2010; Lin et al., 2008), the linkage has some critical problems. First, within the domain of unconstrained econometric or rule-based models, the congestion effect does not optimally change the ATPs; instead, it affects some integrated attributes such as trip generation rate and OD pair values (Miller and Roorda, 2003; Lin et al., 2008b; Konduri et al., 2012). Second, as discussed by Najmi et al. (2018b), the feedback loop is limited to simulation of TPMSs; thus, any error at the estimation of the models may result in a mismatch between the models at the initial stage of simulation which can be problematic in the subsequent iterations. The feedback loop between the models propagates the errors in the TPMSs while the iterations proceed. To rectify the second problem, a possible solution can be calibrating the whole TPMS in the presence of the feedback loop hoping to minimise the mismatch between the demand-side and traffic assignment models. Then, the calibrated model can be used in simulation. This approach is implemented in the current study as well as in Najmi et al. (2018b).

In the next section, we develop a comprehensive ARP-based ATPs generator that is linked to multiple traffic assignment models. The linkage is not limited to simulation; rather, the interaction of the ATPs generator and traffic assignment models are considered in the calibration of parameters.

## 3. Proposed supernetwork

At first, we represent the proposed supernetwork structure using the simple example in Fig. 1. In the proposed supernetwork, nodes are either physical nodes, which are real locations in space (physical network nodes), or activity nodes (e.g., working, shopping and parking), which are never visited for moving on the physical network but represents candidate activities that may be conducted in a physical node. Further, any link is either a travel link, on which a movement can happen on a physical network (e.g., private vehicle road, public transport links), or a virtual link, which never changes the location of a traveller but allows transition between different modes or an activity conduction (e.g., links adjacent to parking and activity nodes).

The ATP in Fig. 1 shows that a traveller leaves home (by private car) to conduct an activity at node 8 for 2 hours, then drives to node 4 to park his/her car and switches to public transport to conduct another activity (for 4 hours) at node 14. Using public transport, the traveller goes back to node 4 to pick up his/her car, and finally to return home. There is only one network for each mode for all travellers. In the figure, the blue links interconnecting home, activity and parking locations represent virtual links.

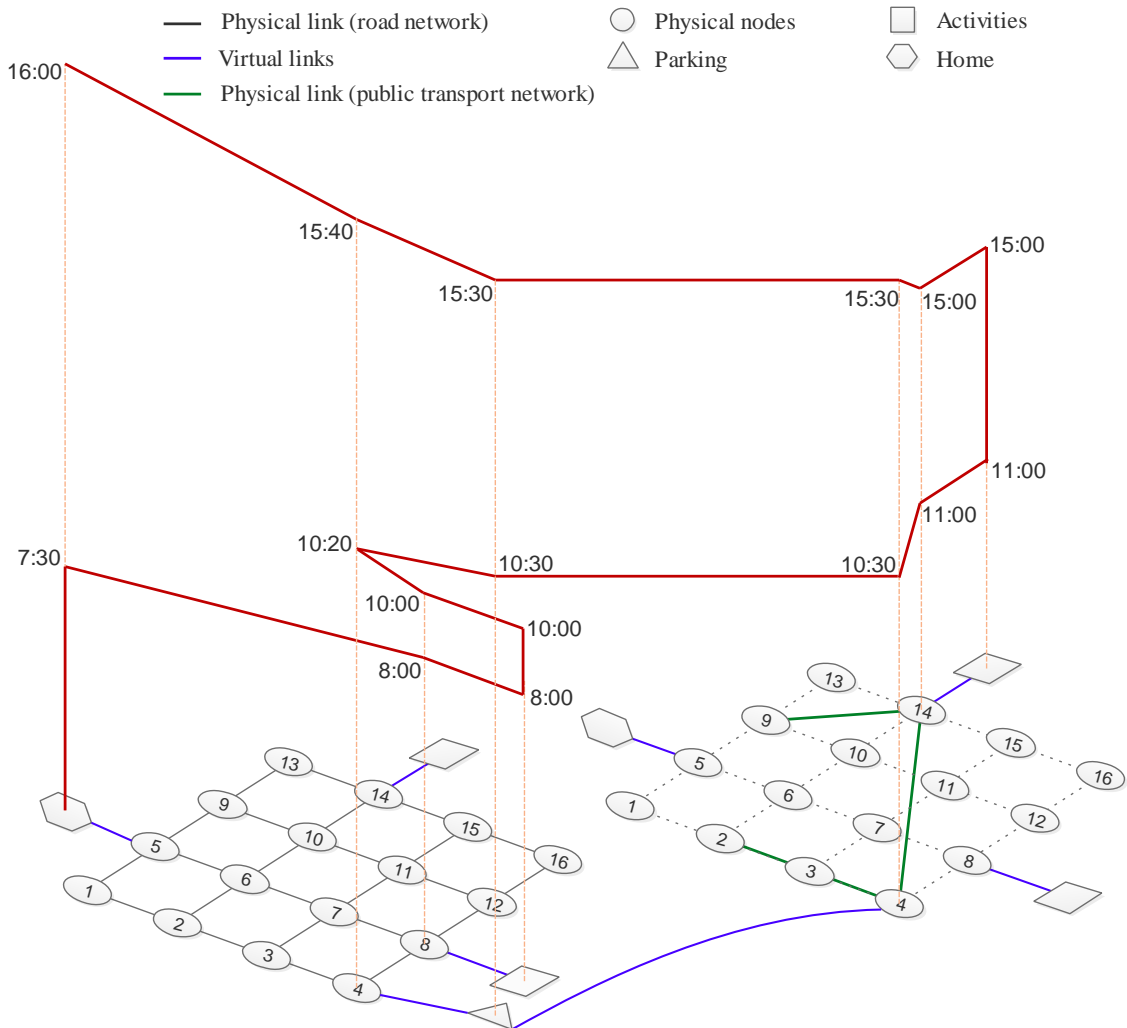

**Figure 1** Supernetwork representation

The problem of visiting a set of activities on a network is an ARP which forms the main structure of the scheduling part of this paper. As it was mentioned earlier, the proposed model of this paper is a generalised ARP with spatiotemporal constraints in which the interaction among travellers is taken into account. The pure ARP is NP-hard in complexity (Lenstra and Rinnooy Kan, 1981), whose extensions make the problem computationally burdensome. Thus, to solve the problem, we propose a pre-processing step, to prepare and scale down the solution space, and a post-processing step to map the solution to the physical network. The whole structure of the proposed model is shown in Fig. 2.

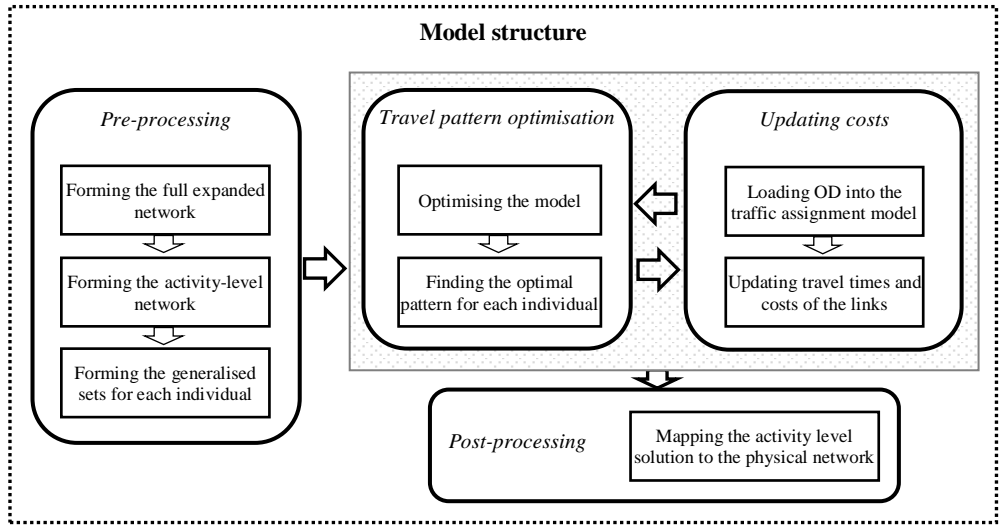

**Figure 2** The main steps of the proposes ATP generation

### 3.1. Pre-processing

Optimising the original connected networks (as in Fig. 1) is quite challenging because 1) while the individual networks (road, transit, and activity networks) within the initial connected networks are spatially connected, there are a number of logical rules and requirements that cannot be addressed by the original network, and 2) the large number of nodes and links within and between the networks represent a substantial computational challenge in solving the problem. To circumvent these obstacles, the steps that should be performed in the pre-processing include: 1) filtering the activity locations, 2) expanding the original connected networks, and 3) forming its mapped network in the activity level.

The ATP of a traveller should reflect where the traveller engages in activities, how and when to get there, and where to park private vehicles (if any) (Liao et al., 2013). The interwoven factors are optimised concurrently when solving the proposed VRP-based optimisation model (see Section 3.2.3) for the scheduling part of the proposed structured. The ideal case is to use all the physical and activity nodes of the network in the optimisation, however, the model is VRP-based and it is computationally burdensome to solve the model with a large number of nodes. Selection of a few nodes for each traveller is an alternative solution to reduce the complexity; however, it may result in undesired ATPs. While considering the trade-off between the size of network and precision of the ATP, designing a location filtering approach is a critical task. One way for tackling the problem of filtering the activity locations is to firstly determine the activity locations (for example based on some logit models) and then to opt for parking nodes based on the activity locations. An interesting filtering approach is provided in (Liao et al., 2013).

To form the expanded network, additional nodes and links should be added to the original connected networks to address the necessary rules in the proposed model. The rules are as follows:

1- A traveller can leave home with a private car, bicycle, public transport, or on foot to conduct out-of-home activities.
2- If a traveller leaves home by private car/bicycle, he/she needs to return his/her home with private car/bicycle at the end of their tour.
3- A traveller cannot switch from the road network to the public transport network unless he/she parks his/her private vehicle in a parking lot at a parking node.
4- If a traveller parks his/her private vehicle in a parking lot, he/she must return to the place to remove the car from the parking lot.
5- If a traveller uses a private vehicle to get to activity locations, he/she either goes to these locations directly or park the vehicle at parking nodes to change his/her modes to get there.
6- A traveller cannot switch to the private vehicle network if he/she has left his/her home by public transport.

Since the home, activity, and parking locations are the main nodes of interest, we map the expanded supernetwork to an activity level to scale down the network size. At the activity level, the cost of each link corresponds to the shortest path on the expanded network. Nonetheless, this reduction does not affect the solution space including the sequences of modes and switching points of the modes as the ATP on expanded the supernetwork could be easily retrieved in a post-processing step to map the activity layer to a physical network.

Fig. 3 shows how a physical network can be converted to an expanded network. In this example, there are two activities that are accessible by public vehicles 1 and 2 and a private car (see Fig. 3a). A traveller can leave home by public vehicle 1 or private car. After reaching to the physical node B, he/she may change mode or continue with the same mode. He/She can move to public transport from the private car network only after parking the private vehicle (see Fig. 3b). As the public transport vehicles have the same rules, all of the corresponding nodes and links for different vehicles are combined and relabelled by PT. An activity should be left with the same mode (PT or private vehicle) as is visited. Then, the activity level can be formed by replacing ineffective physical nodes (or sequences of nodes) with the links. While all the private vehicle nodes can be replaced, only the public transport nodes that are not adjacent to parking nodes and activity nodes are replaced. The adjacent nodes are kept as the travel times on the public transport links not only do depend on the link cost, but also depend on the waiting time at adjacent (boarding) nodes. The cost on the replaced links can be obtained using shortest path algorithms. After pre-processing, the expanded network includes 6 nodes and 11 links (see Fig 3c). Thus, pre-processing allows incorporating the logic rules in the solution space by expanding the network and simplifying the network by neglecting unnecessary information in the network.

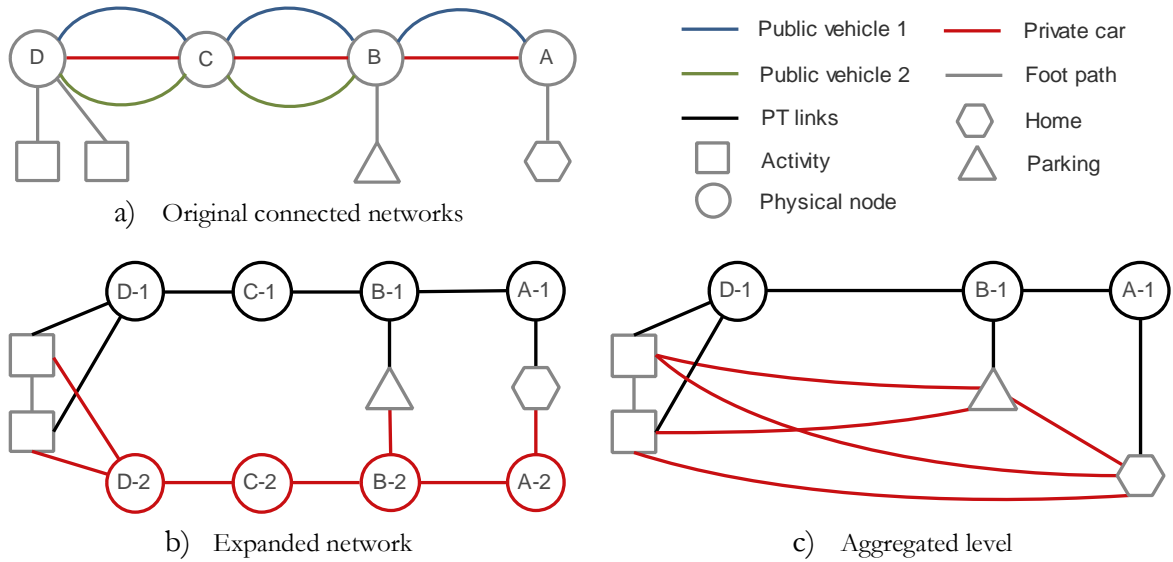

**Figure 3** Expanding network in Pre-processing

### 3.2. Travel pattern optimisation

Generalisation of the ARP and revisiting the nodes are two main concepts that are applied in the model formulation. Thus, before presenting the main body of the model, the concepts are illustrated in the following sections.

#### 3.2.1. Generalised activity routing problem

The Generalised ARP (GARP), which is equivalent to Generalised VRP (GVRP), is an extension of the VRP and was firstly introduced by Ghiani and Improta (2000). The GVRP looks for the optimal delivery or collection routes, subject to capacity restrictions, from a given depot to a number of predefined, mutually exclusive and exhaustive node-sets (clusters) (Pop et al., 2012). The concept of GVRP could be extended to daily activity destination selection where a location should be selected from a set of locations for conducting a specific activity (Kang and Recker, 2013).

Without loss of generality, we assume that each traveller has a set of activity types that needs to be conducted within the spatiotemporal constraints. Some activity types, such as school and work may have only a single candidate destination, whereas each flexible activity such as shopping can be conducted in one of multiple candidate locations. Therefore, no more than one of the candidate locations can be visited. Therefore, the problem is under the category of generalised ARP (Laporte and Nobert, 1983; Noon and Bean, 1991) in which the node set $V$ is partitioned into a number of clusters and the goal is to determine the best cycle, starting and ending at home, and visiting no more than one node in each cluster. Different candidate locations for each activity, assuming the same activity duration, offer the same satisfaction. Fig. 4 depicts the network generalisation on a sample network.

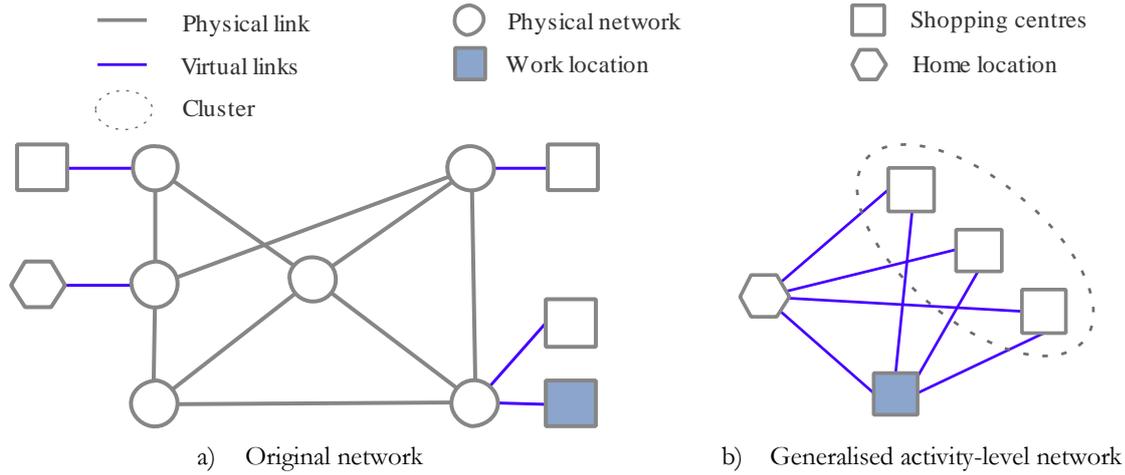

a) Original network        b) Generalised activity-level network

**Figure 4** Network generalization

### 3.2.2. Multi-visit vehicle routing model

Each traveller leaves home to serve a number of activities and then returns home again by the end of the day. Such home-based tours may happen several times in a day for each traveller. In addition, parking locations and work places are other nodes that may be visited multiple times over a scheduling period. Therefore, the problem falls under the category of Split Delivery Vehicle Routing Problem (SDVRP). In the SDVRP, a fleet of capacitated vehicles serve a set of nodes (customers), with known demand, each of which is required to be visited by at least one vehicle (Archetti and Speranza, 2008). In other words, the demand and depot nodes are allowed to be served in more than one visit, thus allowing routes with smaller travel cost.

In the SDVRP literature, to formulate the multi-visit feature, scholars have enumerated either the deliveries to a specific node or the links connected to the nodes. Enumerating deliveries is commonly applied in scheduling problems (see e.g. Berghman et al., 2014; Asbach et al., 2009; Narayanan et al., 2015; Maghrebi et al., 2014). Although both options allow revisiting nodes more than once, enumerating the nodes and link may be costly in the context of supernetworks and VRP-based models where they can significantly increase the solution space. Furthermore, while the traditional revisiting formulations can easily distinguish the routes, they do not provide all scheduling features namely departure time and visiting duration for each revisit. More precisely, expressing the scheduling constraints in the traditional revisit formulations is difficult (Kinable et al., 2014). This is in contrast with the supernetwork and ARP contexts where generating the ATPs for travellers is its main purpose. Thus, a new formulation for the SDVRP is presented in the next section which not only does not enumerate the nodes and links, but even allows tracing the optimal tours, departure time and activity duration in each node revisit.

### 3.2.3. Model formulation

Visiting activities in the space-time prism with the least cost is the main factor that determines the ATP for each traveller. To achieve the optimum ATPs for the travellers, in this section, a mixed integer linear programming problem is presented.

Consider an activity-based supernetwork SNK ($V,E$) composed of a set of nodes, $V$, and a set of directed links, $E$. Let $(i,j) \in E$ denotes a directed link in SNK which connects node $i$ to $j$ and allows the traveller/passenger $p \in P$ to traverse the links to conduct some activities. In addition, the set of home locations of the travellers is represented by $H(p) \in V$. We assume that all the ATPs are home-based in which home is both the origin and the destination of any feasible daily ATP which may include some sub-tours. It should be noted that a traveller may visit its home more than once in an ATP.

The time period [0,T] is chosen such that all possible daily activity-travel patterns are covered. Each node is associated with a departure time window constraint $[\underline{t}_i^p, \bar{t}_i^p]$, where $\underline{t}_i^p$ and $\bar{t}_i^p$ are the earliest and latest time to departure from node $i \in V$ by traveller $p \in P$; and an activity duration $[\underline{d}_i^p, \bar{d}_i^p]$ constraint, where $\underline{d}_i^p$ and $\bar{d}_i^p$ are the minimum and maximum durations that traveller $p \in P$ can spend at node $i \in V$. A traveller may depart a node at different time slots $n \in N$. The departure may happen by different modes of transport $k \in K$. We denote link selection variable $x_{ijll'}^{p,k,n}$, which takes the value of 1 if node $j \in V$ is visited for visit $l' \in L$ immediately after visiting node $i \in V$ for visit $l \in L$ by traveller $p \in P$ and mode $k \in K$ and at time period $n \in N$, and 0 otherwise. The variable determines the type and the sequence of visits to the nodes. We assume that all travellers are heterogeneous so that they should be considered individually; however, they impact each other through congestion happening on the network which depends on their ATPs.

Parameters

| | |
|---|---|
| SNK($V,E$) | supernetwork |
| $i \in V$ | nodes in the supernetwork |
| $k \in K$ | mode of transport which can be categorised into $PT$, $PV$, and $W$ |
| $v \in K(PT)$ | public transport vehicles |
| $S \subset V$ | subset of nodes representing parking locations |
| $i \in V(S)$ | parking nodes in the supernetwork |
| $i \in V(PT)$ | public transport nodes in the supernetwork |
| $i \in V(v)$ | nodes in the supernetwork where public vehicle $v$ has station |
| $(i,j) \in B$ | links in the subtour |
| $y \in Y$ | activity types |
| $i \in y$ | nodes in activity type $y \in Y$ |
| $p \in P$ | traveller $p$ in the supernetwork |
| $H(p)$ | home location of traveller $p \in P$ |
| $n \in N$ | time slots for traffic assignment model |
| $M$ | big constant |
| $l \in L$ | visits |
| $(i,j) \in E(k)$ | links of type $k \in K$ in the supernetwork |
| $u \in U(v)$ | counter of scheduled departure time for public vehicle $v$ |
| $h_{iu}^v$ | $u$th departure time of public transport $v \in K(PT)$ at node $i \in V$ |
| $\tau_{ij}^n$ | travel time on link $(i,j) \in E$ at time slot $n \in N$ |
| $[l_n, u_n]$ | time slot limits for time slot $n \in N$ |
| $[\underline{d}_i^p, \bar{d}_i^p]$ | limits of the spent time on node $i$ by traveller $p \in P$ |
| $[\underline{t}_i^p, \bar{t}_i^p]$ | limits of the departure time from node $i$ by traveller $p \in P$ |

| $\epsilon$ | numerical tolerance parameter |
| --- | --- |
| $\widehat{X}_{yy'}^{PV,n}$ | observed number of trips from activity type $y \in Y$ toward activity type $y' \in Y$ in time slot $n \in N$ by private vehicles |
| $\widehat{X}_{y}^{n,PT}$ | observed number of trips that are originated from activity type $y \in Y$ in time slot $n \in N$ by public transport |
| $\widehat{X}^{n}$ | observed number of trips that are generated in time slot $n \in N$ |

Variables:

| $x_{ijll'}^{p,k,n}$ | 1 if node $j \in V$ is visited for visit $l' \in L$ by mode $k \in K$ immediately after visiting node $i \in V$ for visit $l \in L$ by traveller $p \in P$ and at time slot $n \in N$ and 0 otherwise |
| --- | --- |
| $t_{il}^{p}$ | departure time from node $i \in V$ after visit $l \in L$ by traveller $p \in P$ |
| $d_{il}^{p}$ | time spent at node $i \in V$ by traveller $p \in P$ when visiting the node for visit $l \in L$ |
| $\lambda_{i}^{p}$ | 1 if traveller $p \in P$ visit node $i \in V$ and 0 otherwise |
| $\gamma_{ilu}^{p,v}$ | 1 if traveller $p \in P$ leaves node $i \in V$ after visit $l \in L$ by vehicle $v \in K(PT)$ at its $u$th departure time and 0 otherwise |
| $\beta_{yy'}^{n,PV}$ | proportion of trips that select activity type $y' \in Y$ at time slot $n \in N$ (on the private vehicle network) immediately after conducting activity type $y \in Y$ |
| $\alpha_{y}^{n,PT}$ | proportion of trips that choose public transport (including walking) at time slot $n \in N$ immediately after conducting activity type $y \in Y$ |
| $\delta^{n}$ | proportion of trips that are generated at time slot $n \in N$ |

From here on, the model constraints are discussed. Constraint (1) states that, for each traveller, an ATP must start from home.

$$\sum_{\substack{i=H(p), k \in K, n \in N \\ j:(i,j) \in E(k); l,l' \in L}} x_{ijll'}^{p,k,n} \geq 1 \qquad \forall\, p \in P \tag{1}$$

Constraint (2) is the flow conservation constraint and indicates that, in an ATP, if a node is visited, it must also be left.

$$\sum_{\substack{i:(i,j) \in E(k) \\ n \in N, l' \in L}} x_{ijl'l}^{p,k,n} - \sum_{\substack{i:(j,i) \in E(k) \\ n \in N, l' \in L}} x_{jill'}^{p,k,n} = 0 \qquad \forall\, p \in P,\ \forall\, j \in V,\ \forall\, l \in L, \forall\, k \in K \tag{2}$$

Private vehicle park locations play a key role in supernetworks. They have a transition role to connect the private car and bicycle networks to public transport and pedestrian networks. If people use private vehicles to get to activity locations, either they go to the locations directly, or they park the vehicles at parking nodes to change their modes to get there. The parking nodes are usually at train stations and park-and-ride centres which allow switching to other modes to avoid urban traffic congestion. If a traveller parks their private vehicle in a parking lot, he/she must return to the parking lot to remove the car from it. Since the flow conservation constraint distinguishes between road and non-road links, the parking constraint is already covered by the constraint.

Constraint (3) is a time-window constraint and ensures that a feasible route (sequence of nodes) in the space-time prism will be selected. We denote the travel time on the link $(i,j)$ and time period $n \in N$ by $\tau_{ij}^n$.

$$t_{il}^p + \tau_{ij}^n + d_{jl'}^p - t_{jl'}^p - M\left(1 - x_{ijll'}^{p,k,n}\right) \leq 0 \quad \forall p \in P, \forall k \in K, \quad (3)$$
$$\forall (i,j) \in E(k), \ \forall l, l' \in L, \ \forall n \in N$$

Constraint (4) is the connectivity constraint (sub-tour elimination).

$$\sum_{\substack{k \in K, n \in N, (i,j) \in E(k) \\ i \in B, j \notin B, \ l, l' \in L}} x_{ijl'l}^{p,k,n} \geq 1 \quad \forall p \in P, \ \forall B \subset V \setminus \{H(p)\} \quad (4)$$

where $B$ is a sub-tour formed in the ARP solution. The provided connectivity constraint is an extension of the traditional sub-tour elimination constraint originally developed by Dantzig et al. (1954). In every ARP solution, the constraint forces at least one edge pointing from $B$ to its complement. This means $B$ cannot be disconnected. In this constraint, every node $i \in B$ must be the origin of one edge to another node of $j \in B$ or to a node $j \notin B$.

To form the generalised VRP, $V$ is partitioned into a number of clusters $C^p$. Although there are different clusters of activities in the network, the travellers do not necessarily visit all the clusters. Furthermore, as discussed in section 3.2.2, a node may be visited more than once. Thus, constraints (5)-(7) guarantee that although a node in a cluster can be visited more than once, at most one of the nodes in a cluster is visited.

$$\sum_{\substack{k \in K, (i,j) \in E(k) \\ n \in N; l, l' \in L, j \in V}} x_{ijll'}^{p,k,n} - L\lambda_i^p \leq 0 \quad \forall p \in P, \forall i \in V \quad (5)$$

$$\sum_{i \in \xi} \lambda_i^p \leq 1 \quad \forall p \in P, \forall \xi \in C^p \quad (6)$$

$$\lambda_i^p \in \{0,1\} \quad \forall p \in P, \forall i \in V \quad (7)$$

where $L$ is the maximum number of times that a node may be visited.

onstraints (8) and (9) ensure that departure times and activity durations are properly handled. As each of the activities and facility locations may be revisited several times (an example of which is leaving work place for eating lunch and revisiting the work activity later), the sum of the participation in an activity is accounted as the activity duration which is an important attribute that the travellers are seeking. It should be noted that the time window for departure time and minimum duration of the non-activity nodes are $[0, T]$ and 0, respectively.

$$\underline{t}_i^p \leq t_{il}^p \leq \bar{t}_i^p \quad \forall p \in P, \forall i \in V, \ \forall l \in L \quad (8)$$

$$\underline{d}_i^p \leq \sum_{l \in L} d_{il}^p \leq \bar{d}_i^p \quad \forall p \in P, \forall i \in V \quad (9)$$

Constraints (10) and (11) specify the domain of the decision variables.

$$x_{ijll'}^{p,k,n} \in \{0,1\} \qquad \forall\, p \in P, \forall\, k \in K, \forall\, n \in N, \forall\, (i,j) \in E(k),\ l, l' \in L \tag{10}$$

$$t_{il}^{p}, d_{il}^{p} \in \mathbb{R}_{\geq 0} \qquad \forall\, p \in P, \forall\, i \in V,\ \forall\, l \in L \tag{11}$$

**Public transport links**

Choosing the best route highly depends on the time schedule of public transport vehicles in a transport system. In this section, we abuse the notation by replacing mode $k$ with public transport vehicle $v \in K(PT)$. Each public vehicle $v \in K(PT)$ departs node $i \in V(v)$ for the $u$th visit at time $h_{iu}^{v}$. Furthermore, the binary variable $\gamma_{ilu}^{p,v}$ determines the time when traveller $p \in P$ leaves node $i \in V(v)$ by public vehicle $v \in K(PT)$ at its $u$th visit. Accordingly, constraint (12) determines the $u$th scheduled departure time of vehicle $v \in K(PT)$ that traveller $p \in P$ can get on the vehicle. Specifically, the constraint ensures that a traveller can use a public transport vehicle only if a public transport node is visited.

$$\sum_{\substack{n \in N \\ u \in U}} \gamma_{ilu}^{p,v} - \sum_{\substack{j:(i,j) \in V(PT) \\ l' \in L \\ \in L}} x_{ijll'}^{p,v,n} = 0 \qquad \forall\, p \in P, \forall\, v \in K(PT), \forall\, i \in V(v), \forall\, l \tag{12}$$

Constraint (13) states that the departure time from a public transport node must be equal to a departure time of public transport vehicle. In the equations, the multiplication of scheduled vehicle departure time parameter $h_{iu}^{v}$ to vehicle departure time variable $\gamma_{ilu}^{v}$ determines the time when traveller $p \in P$ departs node $\in V(v)$.

$$\sum_{u \in U, v \in K(PT)} \gamma_{ilu}^{p,v} \left( t_{il}^{p} - h_{iu}^{v} \right) = 0 \qquad \forall\, p \in P, \forall\, i \in V(v), \forall\, l \in L \tag{13}$$

Inequality (14) is the tight vehicle capacity constraint, and constraint (15) defines the public transport usage as an integer variable.

$$\sum_{p \in P, l \in L} \gamma_{ilu}^{p,v} - C_{i}^{v} \leq 0 \qquad \forall\, v \in K(PT), \forall\, u \in U(v), \forall\, i \in V(v) \tag{14}$$

$$\gamma_{ilu}^{p,v} \in \{0,1\} \qquad \begin{array}{l} \forall\, p \in P, \forall\, i \in V(PT), \forall\, v \in K, \\ \forall\, l \in L, \forall\, u \in U(v) \end{array} \tag{15}$$

**Integration of activity-based and traffic assignment models**

Much of the literature on traffic equilibrium models explore the congestion effects in peak period (e.g. Gonzales and Daganzo, 2012; Wahba and Shalaby, 2014; de Cea et al., 2005) which results in ignoring the trip chains and user scheduling behaviour throughout the scheduling period (usually a day) (Chow and Djavadian, 2015). Entering the trip chains into the model has magnified the necessity of having multi-slots (e.g. AM, MD, PM, and EV) in TPMSs with multiple traffic assignment models. The increasing trend of using multi-slots TPMSs (e.g. Miller and Roorda, 2003; Auld et al., 2016) has added to the importance of investigating and improving the interaction behaviour of the multiple traffic assignment models and the multi-slots travel scheduling (demand) models.

Technically, after scheduling the travel demands, the time-dependent ODs are calculated based on the travellers' scheduled trips in each time slot. Then, the ODs are imported into the traffic assignment models which results in the travel time of each link to be updated. While the models are conceptually intertwined, usually their outputs are not consistent in practice. In other words, the OD tables that are consequences of loading travel times into the scheduling models may produce updated travel times that are different from the initially loaded travel times to produce OD tables. It means that the lagged spatiotemporal prism in demand models is not compatible with the time-space prism of the traffic assignment model. To fix this issue, the necessity of an iterative procedure between the travel scheduling model and traffic assignment models is highlighted in the literature (Lin et al., 2008a).

Nevertheless, there is room to optimise/improve the distribution of travel demand across the time slots. To incorporate the influences of traffic assignment models in the ATPs generator, the travel times on the links in each time slot get iteratively updated which necessitates the rescheduling of the ATPs until convergence. This procedure is illustrated in Fig. 5.

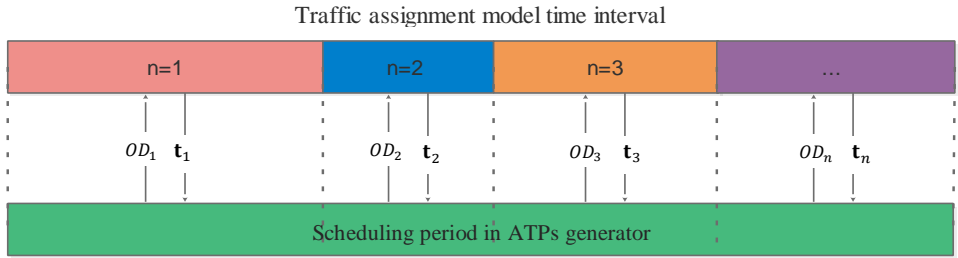

**Figure 5** The interaction between the ATPs generator and traffic assignment model

The start time $l_n$ and closing time $u_n$ of time slots $n \in N$ are applied to partition the scheduling period into some time slots each of which corresponds to a traffic assignment model. Constraints (16) and (17) determine the time slot for the trip departing from node $i \in V$ by traveller $p \in P$. Also, the constraints ensure that each departure time belongs only to one time slot. The constraints affect $\tau_{ij}^n$ in Constraint (3). It is worth mentioning that $\tau_{ij}^n$ is endogenous in the whole structure of TPMS (ARP and TA models) and exogenous to the ARP model.

$$t_{il}^p - u_n - (M - u_1)\left(1 - \sum_{\substack{k \in PV, l' \in L \\ j:(i,j) \in E(k)}} x_{ijll'}^{p,k,n}\right) \leq 0 \qquad \forall\, p \in P, \forall\, i \in V, \tag{16}$$
$$\forall l \in L,\ \forall n \in N$$

$$l_n - t_{il}^p - l_4\left(1 - \sum_{\substack{k \in PV, l' \in L \\ j:(i,j) \in E(k)}} x_{ijll'}^{p,k,n}\right) \leq 0 \qquad \forall\, p \in P, \forall\, i \in V, \tag{17}$$
$$\forall l \in L, \forall n \in N$$

Obviously, there is a trade-off between the number of time slots and the level of accuracy of the integrated model—models with higher number of time slots can capture the reality and congestion on the links more closely, but it demands more computational resources.

**Splitting ratios in the system and calibration constraints**

According to Chow and Recker (2012), there are some characteristics in ARP which make using a utility maximisation approach for parameter estimation challenging. The characteristics are 1) the extremely large number of alternatives, 2) a combination of continuous and discrete variables, 3) non-mutually exclusive choices, and 4) the complexity of the space-time constraint. Because of the complexity of the activity routing problem and also the nature of its solution, their calibration is challenging. To tackle this issue, in Chow and Liu (2012), an inverse optimisation approach is proposed to estimate the coefficients of a set of given objective functions for each traveller. In their calibration solution, the focus is on reproducing the ATPs of each household; therefore, their calibration solution is household level and the calibration of the system level properties is not discussed.

Considering TPMSs' system level properties (such as link travel times among many others) is a critical aspect of the proposed formulation, which albeit indirectly affects travellers' individual level properties. Accordingly, we propose a new solution by introducing splitting ratios in the formulation which distribute trips into the system based on some estimated/calibrated ratios. These ratios control the system level properties the transport network. Thus, the model distributes all generated trips into the network using splitting ratios based on 1) trip purposes on private vehicle network, $\beta_{yy'}^{PV,n}$, 2) time slots, $\delta^n$, and 3) public transport network, $\alpha_y^{PT,n}$. The splitting ratios can be calibrated using the household travel survey data to then use the model for simulation applications. Not only do the splitting ratios control the system level properties of the proposed TPMS, but they also take into account the interaction among travellers. Equations (18)-(25) are the distributing constraints in the model.

$$\sum_{\substack{p\in P, n\in N; l,l'\in L \\ (i,j)\in E(PV), i\in y, j\in y'}} x_{ijll'}^{p,PV,n} - \sum_{\substack{p\in P, n\in N; l,l'\in L \\ j\in y, i:(i,j)\in E(PV)}} \beta_{yy'}^{PV,n} x_{ijll'}^{p,PV,n} - \epsilon \leq 0 \qquad \forall\, y, y' \in Y \tag{18}$$

$$\sum_{\substack{p\in P, n\in N; l,l'\in L \\ i\in y, j:(i,j)\in E(PT)}} x_{ijll'}^{p,PT,n} - \sum_{\substack{p\in P, n\in N; l,l'\in L \\ j\in y, i:(i,j)\in E(k)}} \alpha_y^{PT,n} x_{ijll'}^{p,k,n} - \epsilon \leq 0 \qquad \forall\, y \in Y \tag{19}$$

$$\sum_{y'\in Y, n\in N} \beta_{yy'}^{n,PV} + \alpha_y^{n,PT} = 1 \qquad \forall\, y \in Y \tag{20}$$

$$\sum_{\substack{p\in P, k\in K; l,l'\in L \\ (i,j)\in E(k)}} x_{ijll'}^{p,k,n} - \sum_{\substack{p\in P, k\in K; l,l'\in L \\ n'\in N, (i,j)\in E(k)}} \delta^n x_{ijll'}^{p,k,n'} - \epsilon \leq 0 \qquad \forall\, n \in N \tag{21}$$

$$\sum_{n\in N} \delta^n = 1 \tag{22}$$

$$0 \leq \beta_{yy'}^{PV,n} \leq 1 \qquad \forall\, y, y' \in Y, \forall\, n \in N \qquad (23)$$

$$0 \leq \alpha_y^{PT,n} \leq 1 \qquad \forall\, y \in Y, \forall\, n \in N \qquad (24)$$

$$0 \leq \delta^n \leq 1 \qquad \forall\, n \in N \qquad (25)$$

where $\epsilon$ is the numerical tolerance parameter which is added to equations (18), (19), and (21) to avoid possible infeasibilities. The splitting ratios need to be calibrated before being used for simulation. Thus, there are two distinct models for calibration and simulation purposes whose differences are in the objective function and the distributing equations. While the splitting ratios are estimated parameters and exogenous to the simulation model, they are decision variables and endogenous for the calibration model which means they should be estimated. However, the distributing constraints may not be directly used in the calibration model. The second terms in equations (18), (19), and (21) constitute the multiplication of the decision variables which significantly increases the complexity of the calibration model. Thus, in the calibration process, equations (18)-(25) are replaced with terms (26a)-(26c) in the objective function in which the complicated multiplications do not exist anymore. After solving the calibration model and finding the ATPs at convergence, the splitting ratios are calculated to be used in simulation model.

**Objective function**

A weighted objective function may appear in Eq. (26) as follows.

$$\textbf{min} \quad wZ_c + (1-w)Z \qquad (26)$$

where $Z_c$ and $Z$ may be represented by different or by a combination of objectives. In the calibration step, the objective function includes both $Z_c$ and $Z$ terms; However, $Z$ is the only term in the objective function in simulation step. To illustrate more, while $Z_c$ is to be used in calibration model only, distributing equations of (18)-(25) are just used in simulation model. The objectives for $Z_c$ and $Z$ are shown in Eq. (26a) – (26c) and Eq. (26d) – (26f) respectively.

$$\sum_{y,y' \in Y} \left| \sum_{\substack{p \in P, n \in N; l,l' \in L \\ (i,j) \in E(PV), i \in y, j \in y'}} x_{ijll'}^{p,PV,n} - \sum_{n \in N} \hat{X}_{yy'}^{PV,n} \right| \qquad (26a)$$

$$\sum_{y \in Y} \left| \sum_{\substack{p \in P, n \in N; l,l' \in L \\ i \in y, j:(i,j) \in E(PT)}} x_{ijll'}^{p,PT,n} - \sum_{n \in N} \hat{X}_y^{n,PT} \right| \qquad (26b)$$

$$\sum_{n \in N} \left| \sum_{\substack{p \in P, k \in K; l,l' \in L \\ (i,j) \in E(k)}} x_{ijll'}^{p,k,n} - \hat{X}^n \right| \tag{26c}$$

A very common objective function term is the total travel time over the transport system as given in Eq. (26d).

$$\sum_{\substack{k \in K, (i,j) \in E(k) \\ n \in N; l,l' \in L, p \in P}} \tau_{ij}^n x_{ijll'}^{p,k,n} \tag{26d}$$

A portion of time spent at the nodes is subject to waiting time. Despite, the ideal value for waiting time is 0; however, the time windows on departure times of the activity nodes, time windows on the activity durations, and the scheduled timetable of public transport vehicles may result in undesirable waiting time. For instance, public transport vehicles have their scheduled timetable. Thus, arriving at the station sooner than the scheduled departure time increases the waiting time in the node. For public transport nodes, in particular, the waiting time is equal to the spending time variables $d_{il}^p$ in the nodes. A general formulation for calculating waiting time over all the nodes is given in Eq. (26e).

$$\sum_{\substack{k \in K, (i,j) \in E(k) \\ n \in N; l,l' \in L, p \in P}} \left( t_{jl'}^p - t_{il}^p - d_{jl'}^p - \tau_{ij}^n \right) x_{ijll'}^{p,k,n} \tag{26e}$$

It should be mentioned that Eq. (26f) can be used to keep track of the time spent in parking lots.

$$\sum_{\substack{i: i \in V(S), (i,j) \in E(PV) \\ p \in P, n \in N; l,l' \in L}} x_{ijll'}^{p,PV,n} t_{il}^p - \sum_{\substack{i: i \in V(S), (i,j) \in E(PT) \\ p \in P, n \in N; l,l' \in L}} x_{ijll'}^{p,PT,n} t_{il}^p \tag{26f}$$

### 3.3. Post-processing

In the post-processing, after convergence of the model, the generated ATPs for all the travellers are mapped to the physical network to obtain the exact route of the travellers. While the sequence of activities and the time slots of the departure times are available from the scheduling part of the proposed transport model, the physical routes can be obtained from the traffic assignment models.

## 4. Model calibration and convergence

In this section, we introduce the ideal calibration approach followed by the introduction of an alternative calibration approach to reduce the complexity of calibration. Furthermore, we define a number of convergence criteria each of which can be used as the stopping criterion in the model implementation.

## 4.1. Calibration approach

As it is discussed in Section 3.2.3, the splitting ratios should be calibrated in a calibration model and then be used for simulation. To calibrate the distribution ratios, an ideal calibration structure requires running the ATP with calibration constraints and traffic assignment models iteratively to reach convergence. The calibration structure is depicted in Fig. 6a. This requires a relatively large population size and due to the complexity of the model, it is computationally burdensome. An alternative calibration structure is to iteratively calibrate the splitting ratios using a small sample of population; then, having the calibrated splitting ratios, the simulation of TPMS can be run over the entire population (see Fig. 6b).

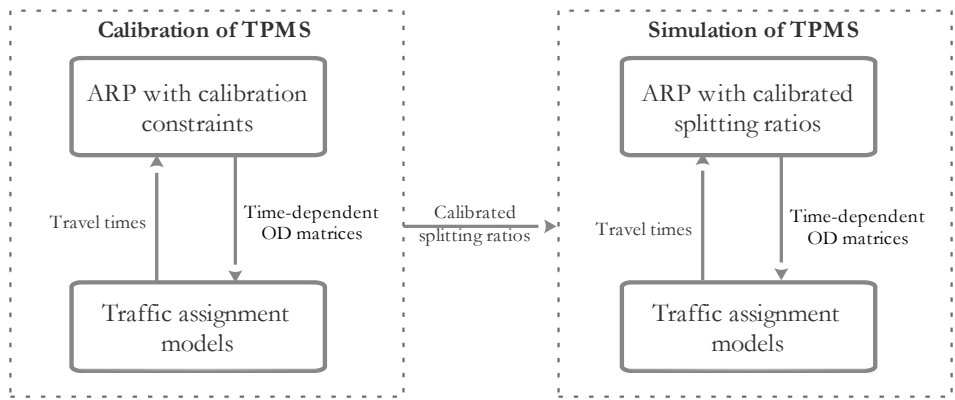

a) Calibration over the whole population

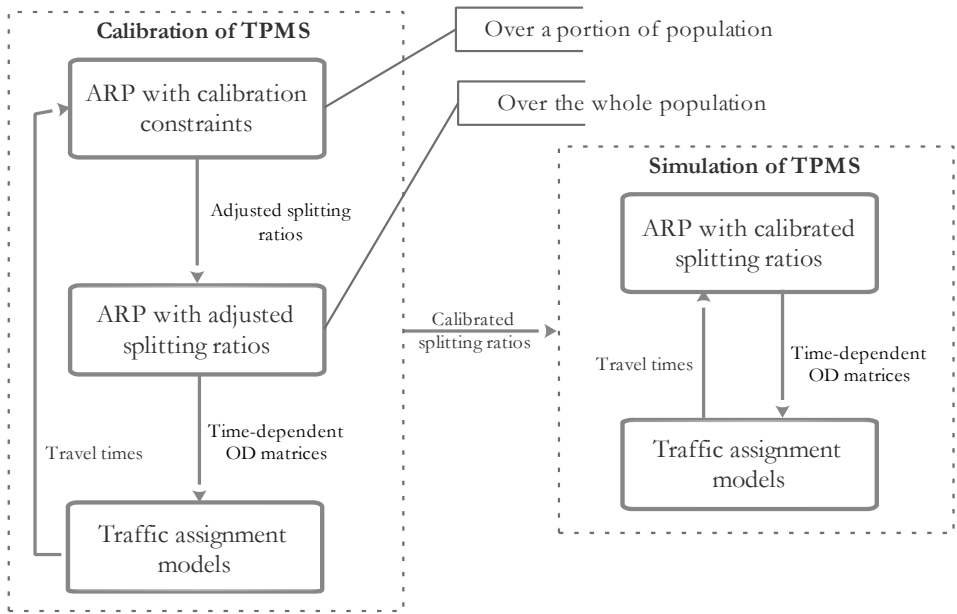

b) Calibration over a portion of population

**Figure 6** Calibration frameworks of TPMSs

A technical problem in the iterative structures is that the updated travel times cannot be directly used in calibration and simulation structures. Updating the ATPs schedules is not straightforward as all the travellers try to re-route and circumvent the congested links. As the travellers selfishly choose their routes purely based on the travel time, these concurrent updates may only shift the congestions to other links which then may even exacerbate the optimality gap in the models. The direct usage of the updated travel times in each iteration usually results in oscillations in performance criteria. The method of successive averages (MSA) (Sheffi, 1985) is used to average the travel times before running the ATP models to prevent an endless cycle of repeated alternation optimum routes in different iterations.

MSA converges to the equilibrium solution in static traffic assignment (STA) problems with well-behaved link-cost functions (Powell and Sheffi, 1982). However, the typically slow convergence rate of MSA in STA (Sheffi, 1985) can be a problem not only in the developed model in this paper but also in large-scale TPMSs in practice, where the computation cost per iteration can be high due to time-space or even solely time expanded routing calculations.

### 4.2. Convergence criteria

As it was shown in Fig. 6, the proposed model iteratively updates OD matrices and travel times. Although travel times and OD values form the main body of the objective function, investigating the convergence behaviour of the model can be fruitful not only with regard to the changes in travel times and OD values but also with regard to the changes in some other criteria such as trip generation rates and splitting ratios. The behaviours can determine the stopping criterion in practice.

Four measures of convergence can be used: 1) OD matrices convergence (ODC), 2) travel time convergence (TTC), 3) trip generation convergence (TGC), and 4) Splitting ratios convergence (SRC). If the average of difference in the subsequent iterations is less than a predefined stopping criterion, the algorithm stops whose output is the converged solution. The convergence criteria are outlined in equations (27a)-(27d):

$$ODC = \sum_{od \in OD} \frac{|OD_{od}^r - OD_{od}^{r-1}|}{OD_{od}^{r-1}} \times 100 \qquad (27a)$$

$$TTC = \sum_{p \in P} \frac{|TT_p^r - TT_p^{r-1}|}{TT_p^{r-1}} \times 100 \qquad (27b)$$

$$SRC = \sum_{y,y' \in Y} \left|\beta_{y,y'}^r - \beta_{y,y'}^{r-1}\right| \qquad (27c)$$

$$TGC = \sum_{n \in N} \frac{|TG_n^r - TG_n^{r-1}|}{TG_n^{r-1}} \qquad (27d)$$

where $OD_{od}^r$, $TT_p^r$, and $TG_n^r$ are the total number of trips between OD pair $od$, total travel cost for traveller $p$, and total number of trips at time slot $n$, at iteration $r$, respectively.

## 5. Model complexity

Finding a feasible solution for the VRP with time window problem in itself is an NP-complete problem (Savelsbergh, 1985). This is the reason that heuristics play a key role in solving the problems. Nevertheless, realistic size instances are solvable optimally through mathematical programming techniques when the problem is sufficiently constrained (Cordeau et al., 2007). As the proposed model in this study is a generalisation version of VRP with time window, and also the calibration constraints are introduced, the model is quite complex to solve. The complexity is indispensable because capturing the determinant rules in real-world activity scheduling of travellers requires an extensive number of variables as well as linear and integer constraints.

As the current study has concentrated on the development of a comprehensive formulation for any common transport system, we do not intend to introduce heuristic methods to solve the problem. Nonetheless, a simple version of the model with a reasonable size (to show the capability of the model) is solved in Section 6 through mathematical programming techniques.

## 6. Numerical example

In this section, we conduct numerical experiments to evaluate the performance of the proposed ATPs generator and its integration with traffic network assignment on the Sioux Falls network.

### 6.1. Data and Simulation

The Sioux Falls network was originally proposed by LeBlanc (1988), based on a simplified road network of Sioux Falls. It originally contains 24 nodes and 76 links. The network's spatial configuration is shown in Fig. 7. Although the link capacities and traffic flows are originally per hour, we modified the lane configuration of the network by changing the scheduling period (turning from one to 14 hours) and adjusting the link capacities. The scheduling period is from 7:00am to 9:00pm which, in some of the conducted experiments, is partitioned into four time slots of 7:00am to 10:00am, 10:00am to 2:00pm, 2:00pm to 5:00pm, and 5:00pm to 9:00pm. The original demand of Sioux Falls network approximates 336,000 veh/h, however, in these numerical experiments, we adjust the travel demand to about 130 veh/h based on the new roadway capacities. To have this number of trips, we generate activity travel patterns for 600 people as synthesised datasets so that each person can have on average 3 trips in his/her activity pattern.

The parameters in Table 1 are used to generate the synthesised datasets. Ten random replications of travellers with randomly selected attributes are generated to be able to accurately compare and assess the performance of different variants of the model. To generate the random streams, we firstly generate the activity type and their sequences that the activity types must be met. We consider 4 activity types of work, shopping, service and education. All nodes on the Sioux Fall network are potential home and work locations. However, as it is depicted in Fig. 7, a limited number of activity spots (2 shopping centres, 3 educational centres and 2 service centres) are chosen and located on the network.

Table 1 Parameters used to generate synthetic population.

| Parameters | Values |
|---|---|
| Alternative specific constant | 5 |
| Travel-time coefficient | -0.1 |
| Number of time slots | 4 |
| Number of home locations | 24 |
| Number of work locations | 24 |
| Number of shopping center locations | 2 |
| Number of service center locations | 2 |
| Number of educational locations | 3 |
| Commuting Probability | 0.7 |
| Going to a service center probability | 0.6 |
| Going to a shopping center probability | 0.8 |
| Going to an educational center probability | 0.5 |
| Time spent at home before leaving | Randomly from [0,350] |
| Work duration | Randomly from [300,540] |
| Service duration | Randomly from [15,120] |
| Shopping duration | Randomly from [15,120] |
| Education duration | Randomly from [240,360] |

After determining the activity types and their sequences, the activity locations are determined. We differentiate between different activities to be met. In these experiments, the home locations for each traveller is determined first, and then the location of fixed activities such as work and school are generated by giving the higher weights to the nodes closer to their home. For this purpose, a simple logit model is used with the alternative specific constants and travel-time coefficient parameters provided in Table 1. After assigning the home and fixed activity locations, the flexible activity locations are generated for each traveller using the same logit model. We assume that either the work trip or the educational one is allowed to be included in an activity travel pattern, if any.

The time window for the departure time and the duration of activities are randomly selected using a uniform distribution from the ranges provided in Table 1. Furthermore, since only four time periods are considered for a day in this study, and the time periods are wide, we intentionally use a relatively large time window for the departure time of the flexible activities such as shopping and service centres to allow them to easily move between time slots.

It should be mentioned that the public transport part of the model is not used in the configuration analysis due to the complexity of the formulation.

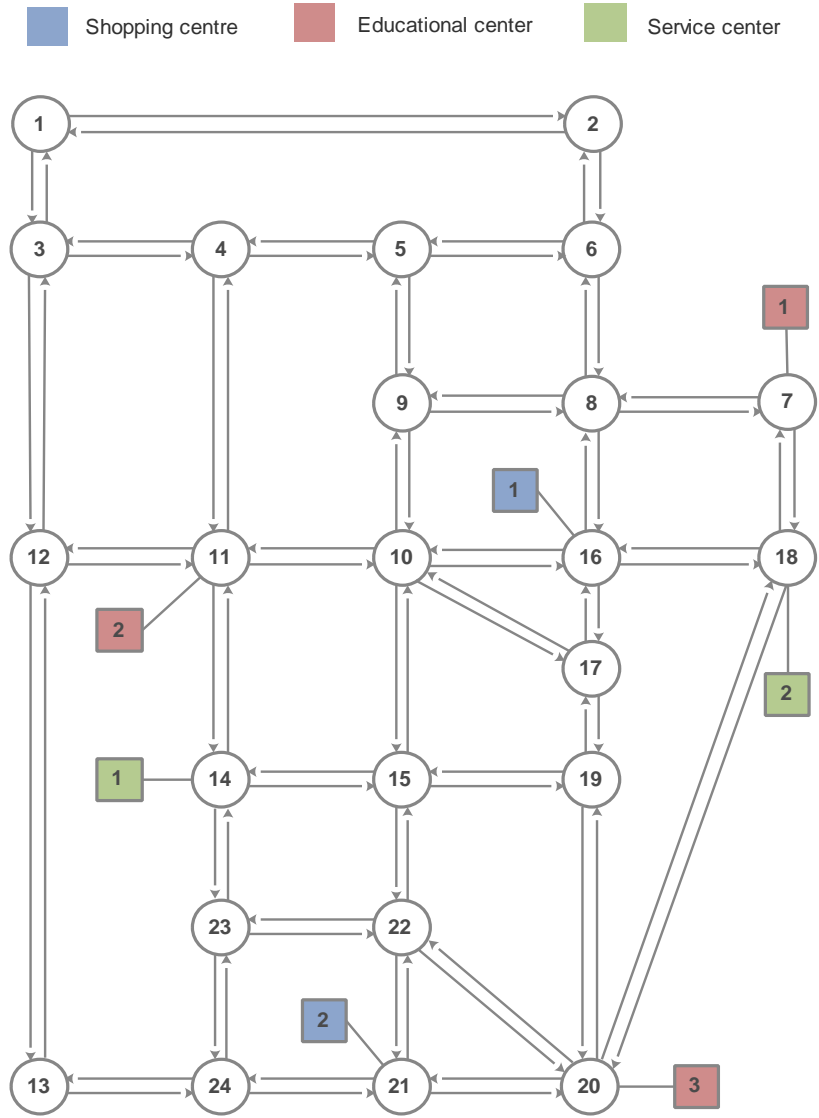
**Figure 7** network activity locations

## 6.2. Model configurations

The goal of the case study is to 1) analyse the convergence of the single-slot versus multiple-slots TPMSs, 2) analyse the performance of splitting ratios in reproducing the observed activity patterns, and 3) compare the performance of the single-slot versus multiple-slots TPMSs. To illustrate the application of the proposed model, there are four variants discussed in this paper. These variants are abbreviated as follows: STWOS (single time slot without splitting ratios), STWS (single time slot with splitting ratios), MTWOS (multiple time slots without splitting ratios) and MTWS (multiple time slot with splitting ratios). In STWOS, there is a single time slot partition for the activity travel pattern problem and consequently a single traffic assignment model. In addition, there is not any calibration model, thus the ARP model without the calibration constraints and network models are iteratively solved. In STWS, the single time slot ARP and single traffic assignment model are iteratively solved in the presence of calibrated splitting ratios. In MTWOS, there are multiple time-lots for ARP and multiple traffic assignment models yet no splitting ratios. In MTWS, there are multiple time-lots for ARP and multiple traffic assignment models in the presence of the calibrated splitting ratios. It is worth to emphasise that the variants with the splitting ratios should be calibrated first to estimate the splitting ratios. In STWOS and MTWOS, the ARP and traffic assignment models are iteratively solved to reach convergence. However, STWS and MTWS include two phases, 1) the ARP for calibration, ARP for simulation, and network models are iteratively solved to find the converged splitting ratios, and 2) the calibrated splitting ratios are used for simulation, which means that the ARP with calibrated splitting ratios and network models are run iteratively to reach convergence.

### 6.3. Simplified calibration model

The alternative structure in Fig. 6b is used to calibrate the model. Furthermore, only equations (18),(22) and (26a) are used for the calibration purposes. Still, solving the ARP with the distributing constraint is complicated; thus, to solve the model with exact methods, we 1) use a proportion of the population (20 percent) to solve the ARP model, 2) ignore the time slot parameter in the splitting ratio and thus a purpose-specific splitting ratio is generated in the calibration model, and 3) consider the entire day as a single-slot and solve a single-slot ARP model which is in line with purpose-specific splitting ratio. Despite the fact that the ARP model for calibration is single slot, the ARP with adjusted splitting ratios is multiple slots in MTWS and MTWOS variants. Thus, the average of the travel times obtained from the network is used in the single-slot ARP model for calibration.

### 6.4. Convergence and evaluation criteria

To investigate the convergence behaviour of the variants, we use the convergence criteria (27a-c). Despite the fact that the convergence criteria can show the performance of the system, we use an activity travel pattern reproduction (ATPR) measure to evaluate the performance of the model in reproducing the observed activity travel patterns. Furthermore, ATPR can be used as another convergence criterion. Thus, Eq. (28) is used to calculate the percentage of the observed activity travel patterns that are generated by the proposed model.

$$ATPR = \left(1 - \frac{\sum_{p \in P; y,y' \in Y} \left| \sum_{\substack{n \in N; l,l' \in L \\ i \in y, j:(i,j) \in E(PV)}} x_{ijll'}^{p,PV,n} - \hat{x}_{y,y'}^{p} \right|}{\sum_{p \in P; y,y' \in Y} \hat{x}_{y,y'}^{p}}\right) \times 100 \qquad (28)$$

All algorithms of the proposed supernetwork-based multi-modal trip chaining models are implemented in Python 2.7 on a machine with 16 GB of RAM with a processor of i7-4770. The optimisation problem is coded in Pyomo (Hart et al., 2011), a free and open-source algebraic modelling language developed in Python, and CPLEX solver is applied for solving the problem. The computational results are presented in the following subsection.

### 6.5. Computational results

Fig. 8 shows the convergence behaviour of different variants. Fig. 8a reveals that the travel times for all the variants converge after a few iterations. It can be observed that the variants experience a steep nose dive in the very first iterations. For STWS and STWOS, the TTC values level out at about zero in the 5[th] and the 10[th] iterations; however, it is the 20[th] iteration that MTWS and MTWOS meet zero. The TTC values for all the variants remain just above 0% after 20[th] iterations.

Figures 8b and 8c depict the performance of the variants in terms of SRC and ODC. Compared to the corresponding STWS, its multiple time slot version of the model (MTWS) converges faster; both in terms of SRC and ODC. This is to some extent expected because partitioning the day allows for 1) more detailed travel time, and 2) higher number of splitting ratios both leading to more accurate predictions and as a result smaller gaps across iterations. The splitting ratios in MTWS converge after a few iterations while the SRC speed for STWS is very slow. Specifically, the SRC for MTWS remains below 0.5 after the 9[th] iteration to reach convergence.

Comparing the behaviour of the variants reveal that the OD matrices that are generated in the variants without the splitting ratios are not converged. This means that the OD matrices in the subsequent iterations are not consistent although the total cost is converged.

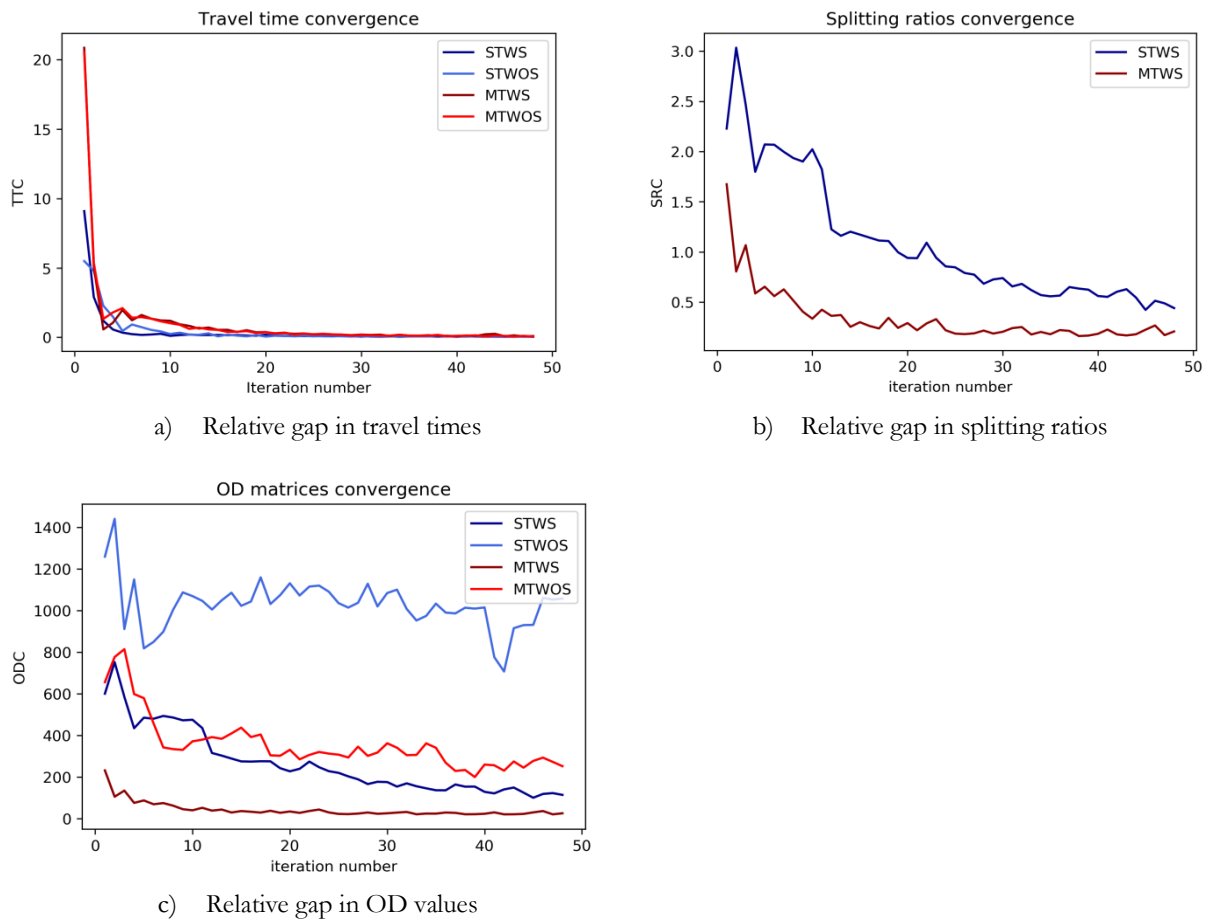

a) Relative gap in travel times
b) Relative gap in splitting ratios
c) Relative gap in OD values

**Figure 8** Evolution of the relative gap for the proposed TPMS variants

Fig. 9 shows that the splitting ratios are absolutely effective in reproducing the observed patterns. The reproduction rates for STWS and MTWS are almost above 50 percent which shows significant improvement in the variants in comparison with STWOS and MTWOS within which the splitting ratios are not included. Also, the figure reveals that STWS and MTWS variants converge in term of ATPR after a few iterations. Although there are a few spikes in the MTWS and STWS cases, they are not significant. An interesting result is that the MTWS converges fast after 8 iterations. Not only do the variants without splitting ratios generate disappointing ATPR rates, but also they hardly converge. This shows the determinant role of splitting ratios in the proposed model.

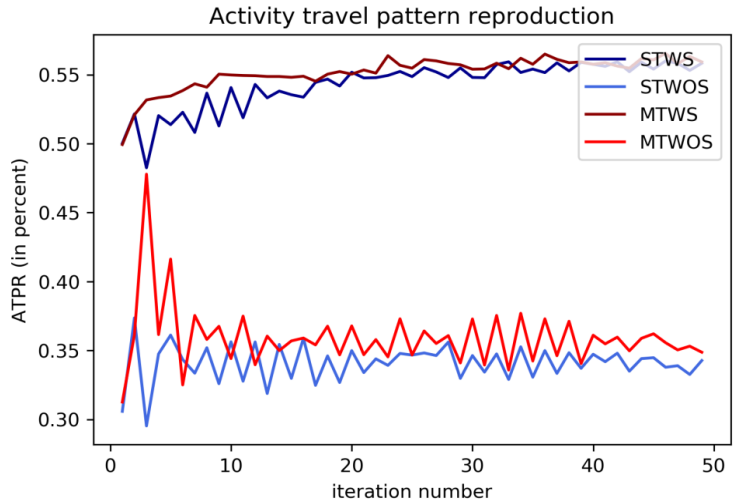

**Figure 9** Reproducing the observed trip patterns (more is better)

Another interesting outcome is the trip generation profile of the multi-slots variants. Fig. 10 presents the importance of the feedback loop in the variants. Before the 8$^{th}$ iteration, the number of trips over different time slots is extremely unstable. Afterward, the models have converged although there are a few spikes in the MTWOS case. Furthermore, trip generation rates for the first iterations represent a condition that feedback loops are not presented in the model. It can be seen that the values for iteration 1 are significantly different from the corresponding rates after convergence (where the feedback loops are included). For example, the number of trips generated by MTWS for AM is 775 which is significantly different from the corresponding number of generated trips in 10$^{th}$ iteration which is 395.

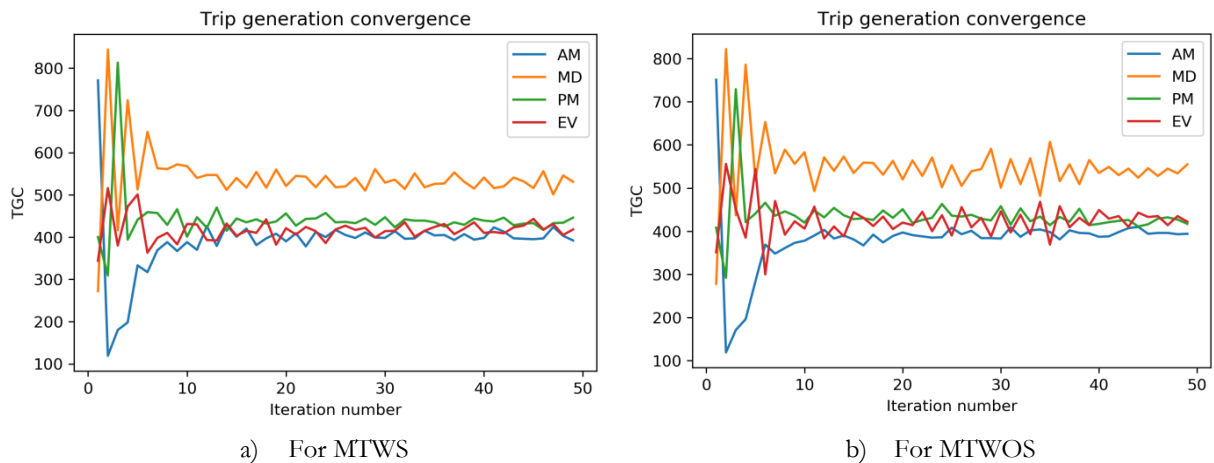

a)  For MTWS                                           b)  For MTWOS

**Figure 10** Trip generation profile

Figures 11 and 12 depict the profiles of different trips of the travellers to get to the location of the activities and the profile of conducting different activities, respectively, over the 14 hours in minutes. In these figures, the plots for the 1st and last iterations are provided; while the models have converged in the last iterations, they have not converged in the 1st iteration. In the plots for MTWS and MTWOS, the vertical lines separate the time slots. Furthermore, the plots for the STWS and STWOS variants are relatively gentler. The reason is that, in MTWS and MTWOS, each travel time varies prior to and after the limit point. It should be highlighted that explaining the shares of the trips and activities in each of the diagrams is not worthy as they are highly sensitive to the random streams (datasets); however, comparing the profiles of the variants across iterations highlights the behaviours of the splitting ratios and feedback loops.

Fig. 11 depicts that while there is a significant difference between the travelling profiles of MTWS and MTWOS at their first and last iterations, the differences are negligible for the STWS and STWOS variants. It discloses the importance and effectiveness of feedback loops in changing the share of trip profiles in multi-slots models. Comparing the variants which include splitting ratios versus those without splitting ratios reveals that the travelling profiles is to some extent insensitive to the existence of the splitting ratios in the multi-slots variants (see the plots for MTWS and MTWOS). Furthermore, the sensitivity is not significant for single-slot variants (see the plots for STWS and STWOS).

Fig. 12 shows the distributions of conducting different activities (the share of time that people spend on activities) over the scheduling period in different variants. The results are to some extent the same as for the Fig. 11. While the splitting ratios are more influential in changing the profiles of activities in single slot variants, the feedback loops are more influential in changing the profile in multi slots ones.

All in all, the experimental results highlights are: 1) the ARP outputs are extremely unstable at the first few iterations which reveal the importance of the existence of the feedback loops in the transport models, 2) using the splitting ratios can be an effective solution to calibrate ARP models, 3) the splitting ratios can speed up achieving the convergence considerably and improve models' performance in reproducing the observed patterns.

## 7. Discussion and Conclusion

*Summary of findings:*

In this paper, we developed a new TPMS which include a unified formulation to obtain the ATPs of travellers in a transport network across multiple dimensions of travel choice while accounting for congestion effects. ATPs generator of the TPMS is an expanded network-based model which is converted to a generalised activity routing problem through pre-processing. In the activity-level representation of the problem, nodes are activities and facility spots (such as parking locations) that are joined by means of travel links. Any route that fulfils the spatiotemporal constraint is a feasible ATP, and the model seeks the optimal ATP (simultaneous determination of activity location, time of participation, duration, and route choice decisions) for each traveller considering the constraints imposed for each traveller. It should be noted that while all the analysis is done at the activity-level, the map of the solutions on the physical network can be obtained by post-processing. Furthermore, in the proposed model, the splitting ratios are used not only to calibrate the model but also to speed up the convergence speed of the model.

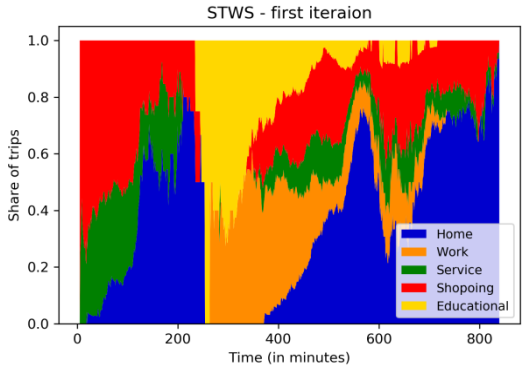
a) Share of trips for STWS (1st iteration)

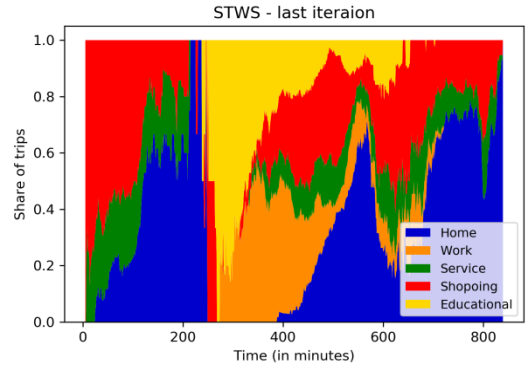
b) Share of trips for STWS (last iteration)

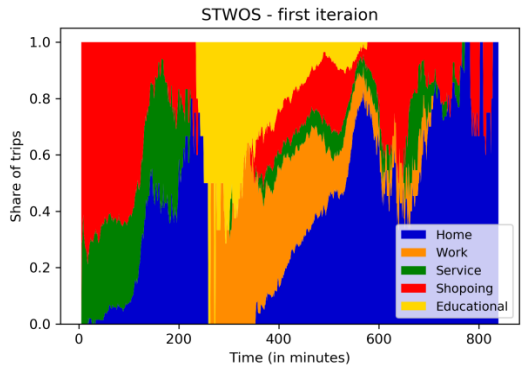
c) Share of trips for STWOS (1st iteration)

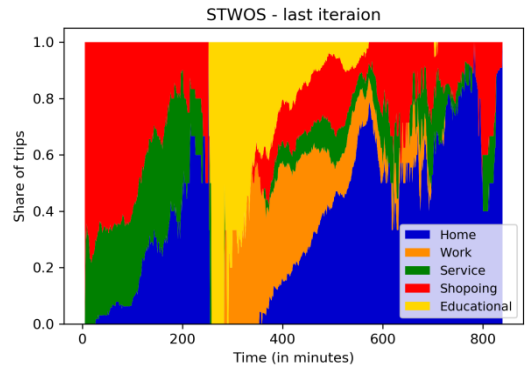
d) Share of trips for STWOS (last iteration)

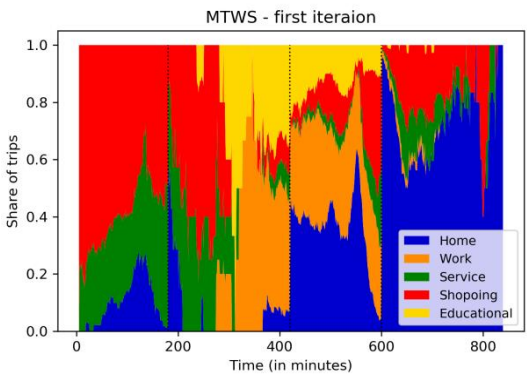
e) Share of trips for MTWS (1st iteration)

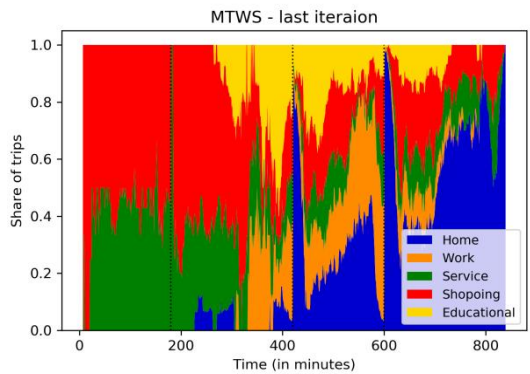
f) Share of trips for MTWS (last iteration)

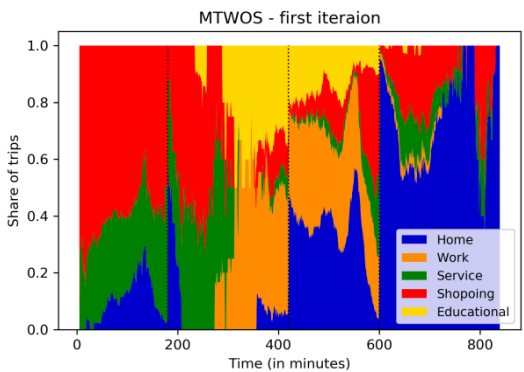
g) Share of trips for MTWOS (1st iteration)

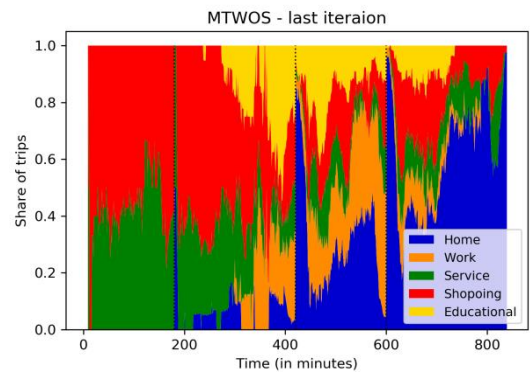
h) Share of trips for MTWOS (last iteration)

**Figure 11** Trip generation profile for different variants

The numerical results highlight the practicality of using the feedback loops and splitting ratios in the numerical experiments which can be potential implications for transport models in practice. While feedback loops are indispensable to reach a convergence in activity routing problems, splitting ratios can remarkably speed up convergence. Specifically, comparing the variants with and without splitting ratios put accent on the following issues: First, the OD matrices that are generated in the variants without the splitting ratios are less similar to observed data. This means that the OD matrices are not consistent in subsequent iterations despite the fact that the total cost is converged. Second, the higher number of splitting ratios, the more SRC and ODC speed up. Third, the splitting ratios converge after a few iterations in MTWS while the SRC speed for STWS is very low.

*Policy implications:*

Most existing approaches fall short of fully representing activity-travel patterns such that some important attributes of transport systems are ignored. To name a few, network congestion, continuum time, and activity sequence generation are the attributes that are, more or less, ignored in different studies. More importantly, the models are usually downgraded from activity travel patterns to trips or to a sequential structure. Furthermore, without taking into account temporal and spatial dimensions of activity locations and without considering their dependencies, the structure of standard transport models, the sequence of activities and also activity locations may be drastically affected. Thus, the model may output inaccurate or even wrong predictions in activity patterns and locations.

We proposed a novel model to address all the travel choice components in a unified structure. The proposed model incorporates spatiotemporal characteristics in activity selection. The proposed model in this paper may be utilised to investigate different scenarios with a higher accuracy. The scenarios include: 1) traffic policies such as road improvement, public transport improvement, adding or removing facility spots such as shopping centres, parking lots and parks, 2) behavioural policies such as the impact of flexible working time on the behaviour of people and the congestion of the roads and the impact of changes in tolling corridors on the activity travel patterns.

*Future directions:*

This study opens up some research directions. First, the developed model involves a non-convex constrained optimization problem and is solvable by exact algorithms only for problems of small sizes. Proposing heuristics for solving the model is highly appreciated. Second, in this paper, we attempted to develop a comprehensive transport model, thus the model is tested on some highly simplified scenarios; nonetheless, analysing more complicated scenarios using the model is left for future research. Despite the fact that the proposed model allows considering the interactions between public transport and private vehicle modes, investigating the interaction would be a challenging stream of research. In addition, analysing and investigating the behaviour of travellers in different scenarios has been left to future research. A general group of scenarios can be considering the impacts of change in network structure on the people' travel behaviour. Third, the proposed model's structure is different from the standard transport models that are being applied in practice. Comparing the performance of these transport models and recognising the performance measures with significant differences can put accent on the shortcomings of standard models. Fourth, the model can be extended to include the timetable synchronisation in public transport as a useful strategy utilised to mitigate the transfer waiting time and as a result improve service connectivity. Because of the multimodality and the tour-based structure of the model, optimising the timetable of public transport may be more realistic.

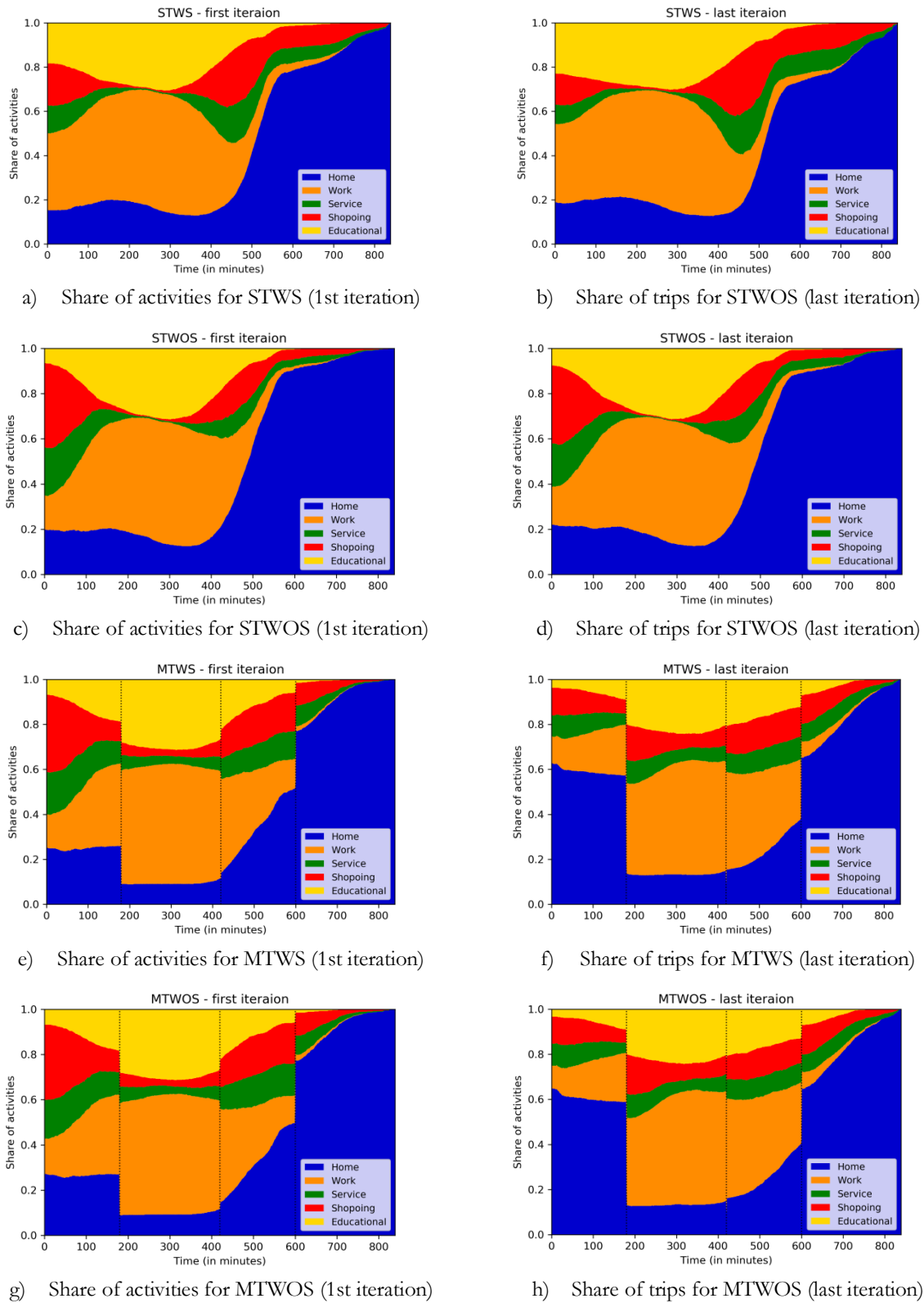

a) Share of activities for STWS (1st iteration)
b) Share of trips for STWOS (last iteration)
c) Share of activities for STWOS (1st iteration)
d) Share of trips for STWOS (last iteration)
e) Share of activities for MTWS (1st iteration)
f) Share of trips for MTWS (last iteration)
g) Share of activities for MTWOS (1st iteration)
h) Share of trips for MTWOS (last iteration)

**Figure 12** Activity profile for different variants

## Acknowledgement

Authors acknowledge the support from the ARC DECRA project (DE170101346).

*This paper has been submitted to Transportation Research part B: Methodological to be reviewed and considered for publication,*